\documentclass[epjST]{svjour}
\usepackage{graphics, amsmath, amssymb, amsfonts}

\begin{document}
\title{Edge state critical behavior of the integer quantum Hall transition}
\author{Martin Puschmann\inst{1}\and Philipp Cain\inst{2} \and Michael Schreiber\inst{2} \and Thomas Vojta\inst{1} }
\institute{Department of Physics, Missouri University of Science and Technology, Rolla, Missouri 65409, USA \and  Institute of Physics, Chemnitz University of Technology, 09107 Chemnitz, Germany}
\abstract{
The integer quantum Hall effect features a paradigmatic quantum phase transition. Despite decades
of work, experimental, numerical, and analytical studies have yet to agree on a unified understanding
of the critical behavior. Based on a numerical Green function approach, we consider the quantum Hall
transition in a microscopic model of non-interacting disordered electrons on a simple square lattice.
In a strip geometry, topologically induced edge states extend along the system rim and undergo
localization-delocalization transitions as function of energy. We investigate the boundary critical
behavior in the lowest Landau band and compare it with a recent tight-binding approach to the bulk
critical behavior [Phys. Rev. B 99, 121301(R) (2019)] as well as other recent studies of the quantum
Hall transition with both open and periodic boundary conditions.
}

\maketitle

\section{Introduction}\label{intro}

Applying a strong perpendicular magnetic field on a two-dimensional free-electron gas leads to highly degenerate eigen energies $E_n=(n+1/2)\hbar\omega$, the Landau levels. Here, $n$ is a non-negative integer and $\omega$ is the cyclotron frequency, $\omega=eB/m$. Disorder lifts the degeneracy and broadens the Landau levels into Landau bands (LBs), leading to extended states in the band center $E_\mathrm{c}$ that separate two localized phases. The integer quantum Hall (IQH) transition is characterized by a power-law divergence of the localization lengths $\xi\sim |E-E_\mathrm{c}|^{-\nu}$ at the critical energy $E_\mathrm{c}$. The value of the localization-length exponent $\nu$ is not settled despite a large body of work in the literature. There are deviations between experimental and theoretical reports as well as between several numerical approaches \cite{LiCT05,SleO09,GruKN17,ZhuWBW19}.

We recently analyzed the IQH transition in a microscopic tight-binding model of non-interacting electrons on a square lattice using the topology of an infinite cylinder \cite{PusCSV19}. By means of a careful scaling analysis, we obtained $\nu=2.58(3)$ in agreement with recent results based on the semi-classical Chalker-Coddington (CC) network model \cite{SleO09,ChaC88,KraOK05,AmaMS11,SleO12,ObuGE12,NudKS15} and other approaches \cite{FulHA11,DahET11}. This value is incompatible with the best experimental results, $\nu\approx2.4$ \cite{LiCT05}.

\begin{figure}
	\includegraphics{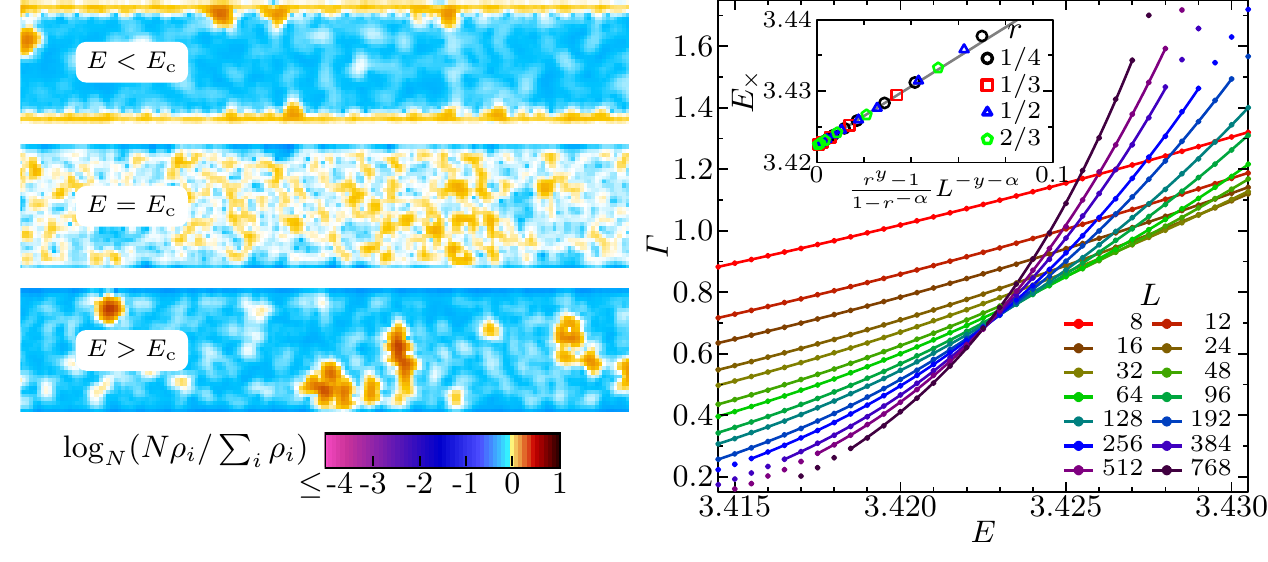}
	\caption{IQH transition in the lowest LB for flux $\Phi=1/10$ and disorder $W=0.5$. Left: Local density of states $\rho_i$ (visualized by color) for a strip of width $L=32$
      for an edge state ($E=3.40$), the critical state ($E=3.42$), and a localized state ($E=3.44$). Only a part of the strip (total length $N=10^4$) is shown.
      Right: Dimensionless Lyapunov exponent $\Gamma(E,L)$ as function of $E$ for several $L$. The statistical errors are well below the symbol size.
      The solid lines are third-order polynomial fits. The inset shows an analysis of the crossing energy $E_\times$ according to Eq.\ (\ref{eq:finitesizecrossing})
      with $y=0.88$ and $\alpha=1/2.6$ for several ratios $r$.}
	\label{fig:Lyapunov}
\end{figure}

In the present work, we make use of the topological features of the IQH effect and consider simple square lattices in a strip geometry with open boundaries. Here, edge states, extended along the system rim, appear, see left panel of Fig.\ \ref{fig:Lyapunov}.
The topological effects change the characteristic of the transition: states above and below the critical energy are localized and extended, respectively, rendering a localization-delocalization transition in the boundary behavior. We study this transition using the recursive Green function method and determine the boundary critical behavior. We find a localization-length exponent $\nu=2.61(2)$ in agreement with the bulk value.

We introduce the model and approach in Sec.\ \ref{sec:model}. Section\ \ref{sec:simulation} is devoted to the analysis of the IQH transition. We conclude in Sec.\ \ref{sec:conclusion}.

\section{Model}\label{sec:model}

We consider a tight-binding model of non-interacting electrons moving on a square lattice of $N\!\times\! L$ sites, given by the Hamiltonian matrix
	\begin{equation}
		\mathbf{H}=\begin{pmatrix}
		\mathbf{H}_{1} & \mathbf{I} &  &  &  \\
		\mathbf{I} & \mathbf{H}_{2} & \mathbf{I} & & \\
		& \mathbf{I}  & \mathbf{H}_{3} & \ddots & \\
		&  & \ddots & \ddots & \mathbf{I}\\
		&  &  & \mathbf{I} &
		\mathbf{H}_{N}
		\end{pmatrix}
		\quad\text{with}\quad\mathbf{H}_x=\begin{pmatrix}
		u_{x,1} & \mathrm{e}^{i \varphi_x} &  &  &  \\
		\mathrm{e}^{-i \varphi_x} & u_{x,2} & \mathrm{e}^{i \varphi_x} & & \\
		& \mathrm{e}^{-i \varphi_x}  & u_{x,3} & \ddots & \\
		&  & \ddots & \ddots & \mathrm{e}^{i \varphi_x}\\
		&  &  & \mathrm{e}^{-i \varphi_x} &
		u_{x,L}
		\end{pmatrix}\;,\label{eq:Hamiltonian}
	\end{equation}
expressed in a Wannier basis. Geometrically, the lattice is a stack of $N$ layers $\mathbf{H}_x$ of $L$ sites each. $\mathbf{H}$ and $\mathbf{H}_x$ have block-tridiagonal and tridiagonal forms, respectively, representing open boundaries (obc) in the $x$ and $y$ directions. The disorder is implemented via independent random potentials $u_{x,y}$, drawn from a uniform distribution in the interval $[-W/2, W/2]$. $W$ characterizes the disorder strength. The hopping terms have unit magnitude, and the uniform out-of-plane magnetic field $B$ is represented via direction-dependent Peierls phases \cite{Pei33,Lutt51}. The hopping in the $y$ direction suffers a complex phase shift $\varphi_x=2\pi\Phi x$ whereas the bonds in the $x$ direction, representing couplings between consecutive layers, do not have phase shifts. This leads to the off-diagonal identity matrices $\mathbf{I}$ in $\mathbf{H}$. $\Phi=Bl^2/\Phi_0$ denotes the magnetic flux through a unit cell (of size $l^2$) in multiples of the flux quantum $\Phi_0=\mathrm{h}/{e}$.

In the clean case, $W=0$, the interplay of the lattice periodicity and the Peierls phases leads to feature-rich Landau-level formation as function of flux $\Phi$, known as the \emph{Hofstadter butterfly}\ \cite{Hof76,Ram85}. In our previous work\ \cite{PusCSV19}, we considered the implications of the butterfly structure (the intrinsic widths and spacings of the Landau levels) for the observation of universal properties of the IQH transition. In particular, we discussed how to chose the  magnetic field values and disorder strengths.
We found that the limit of small $\phi$ represents the best conditions to avoid Landau level coupling.
We then analyzed the bulk IQH transition for $\Phi=1/1000$, $1/100$, $1/20$, $1/10$, $1/5$, $1/4$, and $1/3$ in the lowest Landau band\ \cite{PusCSV19}. We observed universal behavior for $\Phi\lesssim1/10$, where our data collapse when the system size $L$ is expressed in multiples of the magnetic length $L_\mathrm{B}= 1/\sqrt{2\pi\Phi}$. In the current work, we examine the boundary transitions for the same set of system parameters.

We employ the \emph{recursive Green function method} \cite{Mac80,MacK83,Mac85,SchKM84,KraSM84} to characterize the behavior of the electronic states. It recursively computes the Green function $\mathbf{G}(E)=\lim_{\eta\rightarrow 0} \left[(E+i\eta)\mathbf{I}-\mathbf{H}\right]^{-1}$ at energy $E$.
$\mathbf{I}$ is the identity matrix and $\eta$ shifts the energy into the complex plane to avoid singularities. Based on a quasi-one-dimensional lattice with $N\gg L$, the smallest positive Lyapunov exponent,
	\begin{equation}
		\gamma(E,L,\Phi,W)=\lim\limits_{N\rightarrow \infty}\frac{1}{2N} \ln|\mathbf{G}^{N}_{1N}|^2\;,\label{eq:lyapunov}
	\end{equation}
describes the exponential decay of the Green function between the $1$st and $N$th layers. For the current system, the matrix $\mathbf{G}^{N}_{1N}=\mathbf{G}^{1}_{11}\cdot\mathbf{G}^{2}_{22}\cdot\mathbf{G}^{3}_{33}\dots\mathbf{G}^{N}_{NN}$ can be written as product of the diagonal blocks $\mathbf{G}^{x}_{xx}=\left[(E+i\eta)\mathbf{I}-\mathbf{H}_x-\mathbf{G}^{x-1}_{x-1,x-1}\right]^{-1}$. We approximate the limit $\eta\rightarrow 0$ by setting $\eta$ to a small nonzero value, $\eta=10^{-14}$. We use the dimensionless Lyapunov exponent $\Gamma\equiv\langle\gamma\rangle L$ for the scaling analysis. $\langle\gamma\rangle$ represents the ensemble average of $50$ strips of size $L\times10^6$ with width $L$ up to $512$. For $\Phi=1/10$, we improve the accuracy by using $200$ realizations of width $L$ up to $768$.

\section{Simulation and analysis}\label{sec:simulation}

\begin{figure}
		\includegraphics{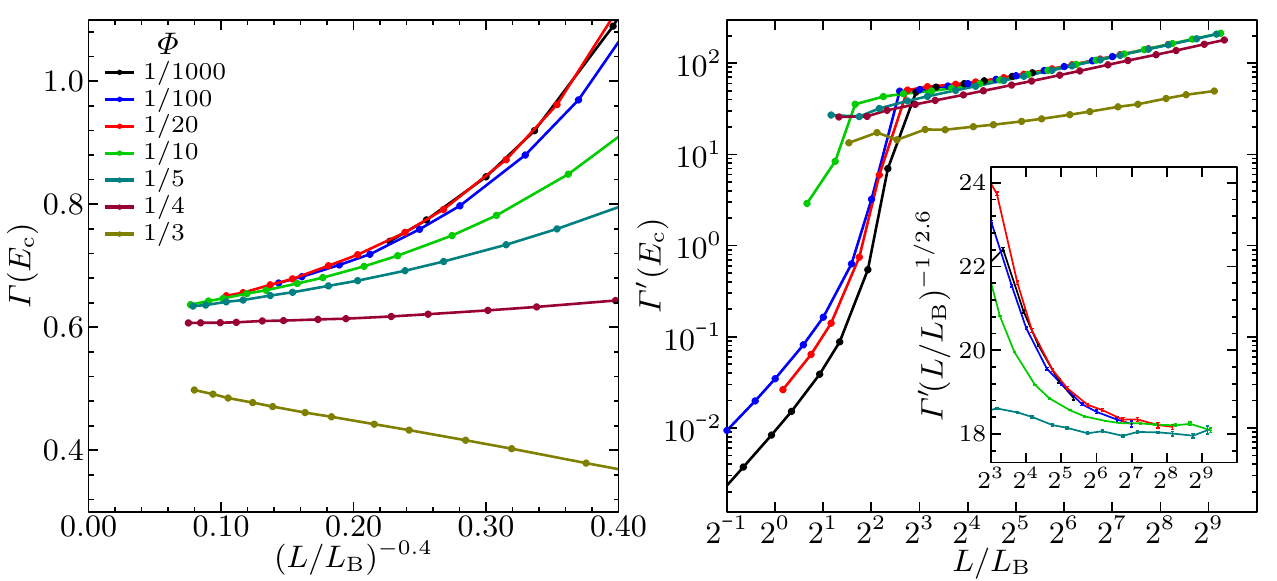}
		\caption{Lyapunov exponent $\Gamma$ and its slope $\Gamma'=\partial\Gamma/\partial E$ at criticality, $E^\mathrm{pbc}_\mathrm{c}$, as function of the effective length $L/L_\mathrm{B}$ for several $\Phi$. Errors are below the symbol size. Lines are guide to the eye only. The inset shows $\Gamma'$ scaled using the relevant exponent $\nu=2.6$, emphasizing that $\nu\approx2.6$ describes the asymptotic behavior for $\Phi\lesssim1/5$.}
		\label{fig:SimpleScaling}
	\end{figure}

Using the recursive Green function method,
we create $\Gamma(E,\Phi,L)$ data sets in the energetic vicinity of the transition in the lowest LB for several $\Phi$. The right panel of Fig.\ \ref{fig:Lyapunov} shows the data for $\Phi=1/10$. We first perform a simple scaling analysis. To this end, we describe the $E$ dependence of $\Gamma$ (for each $L$ and $\Phi$) by a third-order polynomial. For each $\Phi$, we identify $E_\mathrm{c}$ using the crossings of the $\Gamma$ vs. $E$ curves for two different $L$ with ratio $r$, $\Gamma(E_\times,L)=\Gamma(E_\times,L/r)$. The crossings can be extrapolated to infinite $L$ using the scaling ansatz $\Gamma(E,L)= \Gamma_\mathrm{c} + \Gamma_\mathrm{r}(E-E_\mathrm{c})L^{1/\nu} + \Gamma_\mathrm{i}L^{-y}$ with relevant (r) and irrelevant (i) correction terms, which implies
	\begin{equation}
		E_\times(L,r)=E_\mathrm{c}+\frac{\Gamma_\mathrm{i}(r^y-1)}{\Gamma_\mathrm{r}(1-r^{-1/\nu})} L^{-1/\nu-y}\;.\label{eq:finitesizecrossing}
	\end{equation}
The inset of Fig.\ \ref{fig:Lyapunov} shows this extrapolation for $\Phi=1/10$; we use $\nu=2.6$ and $y=0.88$ so that the data for four values of $r$ collapse and the largest number of crossings follow Eq.\ (\ref{eq:finitesizecrossing}), leading to $E^\mathrm{obc}_\mathrm{c}=3.42233(1)$\footnote{We consider fits as reasonable when the mean squared deviation approximates the data's standard deviation. Unless noted otherwise, the given uncertainties of the critical estimates represent statistical standard deviations with respect to individual fits.}.
The fact that the data in the inset nearly perfectly collapse onto the predicted functional form (\ref{eq:finitesizecrossing})
indicates that deviations from the linear energy dependence of $\Gamma$ implied in (\ref{eq:finitesizecrossing}) are not important for crossings of nearby system sizes.
Unfortunately, this extrapolation depends on the (a priori unknown) value of $y$. We can exclude higher values, $y\gtrsim 0.9$ for which the $E_\times$ vs. $L$ curves develop a pronounced S-shape, which would imply that at least three correction-to-scaling terms are important. However, we cannot strictly exclude smaller $y$ values (even though the range of crossings that follow (\ref{eq:finitesizecrossing}) becomes smaller with decreasing $y$). For $y=0.4$, the resulting critical energy, $3.42213(2)$, agrees nearly perfectly with our value $E^\mathrm{pbc}_\mathrm{c}=3.422151(3)$ for the cylinder geometry, where the determination of $E_\mathrm{c}$ is more accurate and robust \cite{PusCSV19}. Within the standard picture of the IQH effect, the critical energies for open and periodic boundaries should coincide in the thermodynamic limit because chiral edge states cannot
Anderson localize due to the absence of back scattering.
This suggests that the above estimate $y\approx0.9$ of the irrelevant exponent is an effective value for our current system sizes only, while the asymptotic value is lower. We perform the same analysis for all $\Phi$; Fig.\ \ref{fig:SimpleScaling} shows the resulting $\Gamma(E^\mathrm{pbc}_\mathrm{c},\Phi,L)$ and their slopes $\Gamma'(E^\mathrm{pbc}_\mathrm{c},\Phi,L)$ at bulk criticality $E^\mathrm{pbc}_\mathrm{c}$.
The data for $\Phi=1/3$ and $1/4$ behave clearly differently from those for lower $\Phi$, whose data asymptotically collapse as function of $L/L_\mathrm{B}$. As in the case of the cylinder geometry \cite{PusCSV19}, we thus consider systems with $\Phi\lesssim 1/10$ to be in the universal regime. If we use $E^\mathrm{obc}_\mathrm{c}$ instead of $E^\mathrm{pbc}_\mathrm{c}$ in Fig.\ \ref{fig:SimpleScaling}, the data collapse is of significant lower quality.

In the following, we use the data for $\Phi=1/10$ for which we have better statistics and larger sizes to extract estimates of the critical exponents and amplitudes. We perform fits at both $E^\mathrm{obc}_\mathrm{c}$ and  $E^\mathrm{pbc}_\mathrm{c}$ to capture errors stemming from the uncertainties of $E_\mathrm{c}$. For $E^\mathrm{obc}_\mathrm{c}$, power-law corrections $\Gamma(E_\mathrm{c},L)=\Gamma_\mathrm{c}(1+aL^{-y})$ lead to reasonable fits for $L\geq32$, yielding $\Gamma_\mathrm{c}=0.6577(3)$ and $y=0.951(7)$. For $E^\mathrm{pbc}_\mathrm{c}$, the simple power-law description is limited to a smaller $L$ range. We obtain $\Gamma_\mathrm{c}=0.614(4)$ with $y=0.49(5)$ and $\Gamma_\mathrm{c}=0.606(13)$ with $y=0.42(14)$ for $L\geq128$ and $L\geq256$, respectively.

We now consider the slope $\Gamma'$ to get estimates for the relevant exponent $\nu$. For both $E_\mathrm{c}$ estimates, we get good-quality fits $\Gamma'(E_\mathrm{c},L)=\Gamma'_\mathrm{c}L^{1/\nu}$ even without irrelevant scaling corrections for $L\geq128$, leading to $\nu=2.556(5)$ for $E^\mathrm{obc}_\mathrm{c}$ and $\nu=2.619(7)$ for $E^\mathrm{pbc}_\mathrm{c}$. For a wider range, $L\geq32$, power-law corrections to scaling need to be included, $\Gamma'(E_\mathrm{c},L)=\Gamma'_\mathrm{c}(1+aL^{-y})L^{1/\nu}$. This yields $\nu=2.523(16)$ with $y=1.2(2)$ and $\nu=2.598(16)$ with $y=1.8(3)$ for $E^\mathrm{obc}_\mathrm{c}$ and $E^\mathrm{pbc}_\mathrm{c}$, respectively.

In addition to the simple scaling analysis, we also perform fits of sophisticated scaling functions $\Gamma(x_\mathrm{r}L^{1/\nu},x_\mathrm{i}L^{-y})$, expanded in terms of relevant and irrelevant scaling field, $x_\mathrm{r}L^{1/\nu}$ and $x_\mathrm{i}L^{-y}$ \cite{SleO99a}. We consider a large collection of such fits based on various subsets of the data and different fit expansions. The results of these fits show fluctuations similar to the results presented above. Hence, whereas the compact fits give robust estimates of $\nu$, they do not give a reliable estimate of $y$, systematically affecting $E_\mathrm{c}$, and $\Gamma_\mathrm{c}$.

\section{Conclusion}\label{sec:conclusion}

In summary, we have investigated the IQH transition in the lowest Landau band in a strip geometry with open boundary conditions for a microscopic model of electrons. In contrast to cylindrical systems, edge states lead to a transition between an extended and a localized phase. Table\ \ref{tab:criticalparams} compares the critical parameters of the IQH transition for the tight-binding model and the CC model for both cylinder and strip geometries.
\begin{table}
	\caption{Critical parameters obtained by means of the tight-binding lattice (TBL) and the CC network model (CCNM) for systems in topology of a cylinder (pbc) and a strip (obc).}
	\label{tab:criticalparams}
	\centering
	\begin{tabular}{l|cc|cc}
		\hline\noalign{\smallskip}
		& TBL, pbc\;\cite{PusCSV19} & CCNM, pbc\;\cite{SleO09} & TBL, obc (current) & CCNM, obc\;\cite{ObuSF10} \\
		\noalign{\smallskip}\hline\noalign{\smallskip}
		$\nu$ & 2.58(3)& 2.593 [2.587,2.598] & 2.61(2) & 2.55(1)\\
		$y$ & 0.35(4)& 0.17 [0.14,0.21] & $\lesssim 0.9$ & 1.29(4)\\
		$\Gamma_\mathrm{c}$ & 0.815(8)& 0.780 [0.767,0.788] & 0.61(1) & 0.6158(8)\\
		\noalign{\smallskip}\hline
	\end{tabular}
\end{table}
Interestingly, literature values of the irrelevant exponent $y$ seem to have a strong dependence on the geometry. Whereas $y$ is very small in cylinders, it is significantly higher ($y\gtrsim 1$) for strips. Does this imply that strong but shorter-ranged boundary corrections are dominant at the current system sizes whereas longer-ranged bulk corrections dominate asymptotically, or do bulk corrections vanish in strip geometry? In the current model, the estimate of $y$ is strongly correlated with the critical energy; a straightforward analysis yields a critical energy marginally different from the bulk value as well as a larger $y$. However, assuming the bulk critical value to be valid, we observe a significant better agreement of $\Gamma_\mathrm{c}$ with the result of the open-boundary CC model investigation.

The main message of the present paper is, however, that the estimate of the localization length exponent $\nu$ is very robust. Combining statistical and systematic errors, we estimate $\nu=2.61(2)$ based on the bulk critical energy, which agrees well with recent high-accuracy CC model calculations. Even, if we consider the variations between different fits combined with the uncertainty of the critical point, we observe $\nu=2.58(5)$, a value considerably different from the experimental value $\nu\approx 2.4$.

\begin{acknowledgement}
This work was supported by the NSF under Grant Nos. DMR-1506152 and DMR-1828489.
\end{acknowledgement}

\noindent M.\,P. and T.\,V. conceived the presented idea. M.\,P. performed the simulations, analyzed the data, and took the lead in writing the manuscript. All authors discussed the results and provided critical feedback to the analysis and the manuscript.

\end{document}